\documentclass[fleqn,usenatbib]{mnras}
\usepackage[T1]{fontenc}
\usepackage{ae,aecompl}
\usepackage{here}
\usepackage{graphicx}	% Including figure files
\usepackage{amsmath}	% Advanced maths commands
\usepackage{amssymb}	% Extra maths symbols

\usepackage{subcaption}
\title[Circular polarization of GWs from non-rotating CCSNe]{Circular polarization of gravitational waves from non-rotating supernova cores: a new probe into the pre-explosion 
hydrodynamics}
\author[Hayama, Kuroda, Kotake, \& Takiwaki]{
Kazuhiro Hayama$^{1}$,
Takami Kuroda$^{2}$,
Kei Kotake$^{3}$, and Tomoya Takiwaki$^{4}$
\\
$^1$ KAGRA Observatory, Institute for Cosmic Ray Research,
University of Tokyo, 238 Higashi Mozumi, Kamioka, Hida, Gifu 506-1205, Japan\\
$^2$ Institut f{\"u}r Kernphysik, Technische Universit{\"a}t Darmstadt
Schlossgartenstrasse 9, D-64289 Darmstadt, Germany\\
$^3$ Department of Applied Physics, Fukuoka University, Jonan, Nanakuma, Fukuoka 814-0180, Japan\\
$^4$ Division of Theoretical Astronomy, National Astronomical Observatory of
 Japan (NAOJ), 2-21-1, Osawa, Mitaka, Tokyo, 181-8588, Japan}
\date{\today}

\pubyear{2018}

\begin{document}
\label{firstpage}
\pagerange{\pageref{firstpage}--\pageref{lastpage}}
\maketitle

  \begin{abstract}
We present an analysis of the circular polarization of gravitational-waves (GWs)
using results from three-dimensional (3D), general relativistic (GR)
 core-collapse simulations of a non-rotating $15 M_{\odot}$ star.
For the signal detection,
we perform a coherent network analysis taking into account 
the four interferometers of LIGO Hanford, LIGO Livingston, VIRGO, and KAGRA.
 We focus on the Stokes $V$ parameter, which directly characterizes the 
asymmetry of the GW circular polarization.
 We find that the amplitude of the GW polarization becomes bigger for 
 our 3D-GR model that exhibits strong activity of the standing accretion shock 
instability (SASI). 
 Our results suggest that the SASI-induced accretion flows to the 
 proto-neutron star (PNS) 
lead to a characteristic, low-frequency 
modulation (100 $\sim$ 200 Hz) in both the waveform and the GW circular polarization. During the vigorous SASI phase, we observe that the GW polarization 
switches from the right- to left-handed mode, which is clearly visible in the 
 spectrogram.
 By estimating the signal-to-noise ratio of the GW polarization, 
we demonstrate that the 
detection horizon of the circular polarization extends by more than a 
factor of several times farther comparing to that of the GW amplitude. 
 Our results suggest that the GW circular polarization, if detected, could
provide a new probe into the pre-explosion hydrodynamics
 such as the SASI activity and the $g$-mode oscillation of the PNS.
  \end{abstract}
 \begin{keywords}
  stars: interiors -- stars: massive -- supernovae: general.
 \end{keywords}

%%%%%%%%%%%%%%%%%%%%%%%%%%%%%%%%%%%%%%%%%%%%%%%%%%

%%%%%%%%%%%%%%%%% BODY OF PAPER %%%%%%%%%%%%%%%%%%

\section{Introduction}\label{sec1}
The LIGO and Virgo collaboration has made the first joint detection 
of gravitational waves (GWs) from merging binary system of black holes 
 \citep{GW2_virgo,GW_review}. The network of the {\it three} detectors has not only
 enabled to improve the sky localization of the source significantly, but also 
 probe the GW polarization for the first time \citep{GW2_virgo}, 
the latter of which was difficult 
 for the twin LIGO detectors having similar orientations. 
With KAGRA \citep{kagra17}, the four-detector era is coming soon.
  The network of these advanced GW detectors is expected to unravel the yet-uncertain
 nature of astrophysical sources (e.g., \cite{schutz09}),
 which include core-collapse supernovae 
(CCSNe, e.g., \citet{janka17} for a review).

Extensive numerical simulations have been done so far to study the 
GW signatures from
 CCSNe in different contexts 
(e.g., \citet{CerdaDuran13,Ott13,Yakunin15,KurodaT14,viktoriya18} and
 \citet{Kotake13,Ott09} for a review). 
For canonical supernova progenitors \citep{Heger05}, core rotation is generally
 too slow to affect the dynamics (e.g., \citet{takiwaki16,summa17}).
For such progenitors, the GW signatures
in the postbounce phase are characterized by prompt convection, neutrino-driven convection, proto-neutron star (PNS) 
convection, the standing accretion shock instability (SASI), and the $g$(/$f$)-mode 
oscillation of the PNS
 surface (e.g., \citet{EMuller97,EMuller04,Murphy09,Kotake09,BMuller13,viktoriya18}).
 Among them, the most distinct GW emission process generically seen in recent 
self-consistent three-dimensional (3D) models 
 is the one from the PNS oscillation \citep{KurodaT16ApJL,Andresen17,Yakunin17}.
 The characteristic GW frequency increases almost monotonically 
with time due to an accumulating accretion to the PNS, which ranges 
approximately from $f_{\rm PNS} \sim 100$ to $1000$ Hz.
On the other hand, the typical 
frequency of the SASI-induced GW signals is concentrated in 
 the lower frequency range of $f_{\rm SASI}\, \sim 100$ to $250$ Hz and persists when the SASI 
dominates over neutrino-driven convection \citep{KurodaT16ApJL,Andresen17}.
%The SASI-induced GW frequency $f_{\rm SASI}\sim100$ Hz is significantly lower 
% than that of the $g$-mode frequency ($f_{g}\sim500$-1000 Hz). 
 The detection of these distinct GW features could 
 help infer which one is more dominant in the supernova engine, neutrino-driven 
convection or the SASI \citep{Andresen17}. In order to discuss the detectability
 of the signals, these studies have traditionally relied on the GW spectrum or 
spectrogram analysis using the information of the GW amplitude and the frequency
 only.

% \citet{Andresen17} pointed out that a third-generation detector like 
% Einstein Telescope \citep{punturo} could 
%distinguish SASI- from convection-dominated case among their full-scale 3D models 
% at a distance of $\sim$ 10 kpc.
%accretion to the PNS. 
% For progenitors with high-compactness \cite{evanott}, the SASI 
% is more likely to dominate over neutrino-driven convection 
% in the accretion phase \cite{Hanke13,Nakamura15}. In such a case, 
%large-scale anisotropic flow associated with the SASI leads to strong GW 
%emission, whose typical 
%GW frequency closely matches with that of the SASI motion \cite{KKT2016,Andresen16}.

 \citet{Hayama16} were the first to point out the importance of detecting
circular polarization of the GWs from CCSNe. They studied the 
GW polarization using results from \citet{KurodaT14}, where only a short duration
 ($\sim 30$ ms) after bounce was followed in 3D general-relativistic (GR) models
of a $15 M_{\odot}$ star, where the initial rotation rates were added to 
 the non-rotating progenitor in a 
 parametric manner. It was found that the clear signature of 
 the GW polarization appears only for their most rapidly rotating 
 model (assuming the initial angular velocity of $\pi$ rad/s in the core).
 While novel, this finding may 
raise several questions. These include whether the GW polarization could or could 
not be generated from canonical (essentially, non-rotating)  progenitors,
how far the GW polarization, if generated, 
would be detectable in the four-detector era, and what we can learn about 
the supernova engine from the future detection of the GW polarization.

 In this work, we aim to answer 
these questions by studying the GW circular polarization using results from 
  3D full GR core-collapse simulations following $\sim 300$ ms after bounce
 \citep{KurodaT16ApJL}.
For the signal detection, we perform a coherent network analysis where 
the network of LIGO Hanford, LIGO Livingston (L), VIRGO (V), and KAGRA (K)
is considered \citep{Hayama15}. 
 We find that the amplitude of the GW circular polarization 
becomes much bigger for our 3D-GR model that exhibits a strong SASI
 activity. 
%This is because the SASI-induced 
%large-scale anisotropic flows striking the PNS surface leads
% to a characteristic low-frequency modulation 
%in both the waveform and the GW circular polarization.
% It is firstly observed that the direction of the 
%circular polarization switches from the right- to left-handed mode 
% in the pre-explosion phase when the SASI activity is vigorous.
%By comparing the signal-to-noise ratio of the GW amplitude 
%with that of the 
% GW polarization, we point out that the detection horizon of the GW polarization
%  extends farther by a factor of several times comparing to that of the 
%GW amplitude. 
 We will show that the GW circular polarization, if detected, could
provide a new probe to decipher the inner-working of the supernova engine.

%show that 
%the $g$-mode oscillation of the PNS core also leads to the emission of the CP
 
%These
% in particular for models that the SASI activity is vigorous ?
%that we shortly call as "CP" for bravity 

\begin{figure*}
\begin{center}
\includegraphics[width=0.9\linewidth]{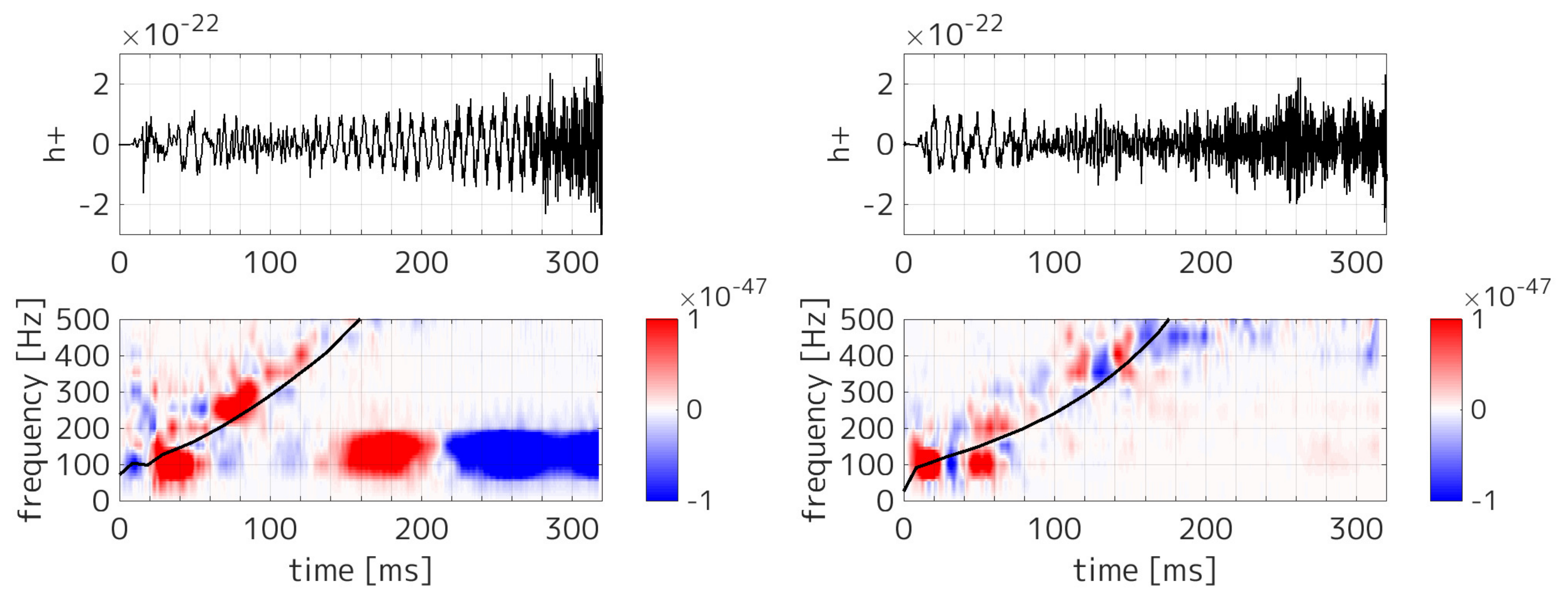}
\caption{The top panels are the original gravitational waveforms ($h_+$) and 
 the bottom panels are the spectrograms for the Stokes $V$ parameter 
 of SFHx (left panels) and TM1 (right panels), respectively. The black line 
 in the bottom panels corresponds to the peak GW frequency of the PNS $g$-mode 
 oscillation (see also Figure 1 of \citet{KurodaT16ApJL}). A source distance 
of $D = 10$ kpc is assumed. The waveforms are extracted along the 
north pole $(\theta,\phi)=(0,0)$. A series of short-time (20 ms) Fourier
 transformation are calculated to obtain the $V$ parameter from the reconstructed 
 waveforms at different times, which we refer to as the spectrogram.
% The sampling rate of them is $16384$Hz. The time and frequency resolution of the distribution are $20$ms overlapped with $19$ms and $8$Hz, respectively.  In the top panels, we 
}
\label{f1}
\end{center}
\end{figure*}

\section{Numerical Methods and Initial Models}\label{sec2}

We analyse the GW predictions from 3D-GR supernova models in 
\citet{KurodaT16ApJL} who followed the hydrodynamics from the onset of 
core-collapse of a $15M_\odot$ star \citep{WW95}, through core bounce, up to $\sim$ 
300 ms after bounce. In the simulation,
 the BSSN formalism is employed to evolve the metric 
\citep{Shibata95,Baumgarte99}, and the GR neutrino transport is solved 
by an energy-integrated  M1 scheme \citep{KurodaT12}.
The 3D computational domain is a cubic box with 15,000 km where 
the nested boxes with 8 refinement levels are embedded.
Each box contains $128^3$ cells and the minimum grid size near the origin 
is $\Delta x=458$ m. 
%In the vicinity of the stalled shock at a radius of
% $\sim100$ km, our resolution achieves $\Delta x\sim 1.9$ km, i.e., 
%the effective angular resolution becomes $\sim1^\circ$. 

 We choose two representative models from \citet{KurodaT16ApJL}, 
 one with the nuclear equation of state (EOS) of SFHx \citep{SFH} and another 
of TM1 by \citet{HS}. 
 In the following, we refer to the two models as SFHx and TM1, respectively.
For SFHx and TM1, the maximum gravitational mass ($M_{\rm max}$) and 
the radius ($\bar{R}$) of a cold NS is
 $M_{\rm max}=2.13$ and $2.21$ $M_\odot$, and $\bar{R}= 12.0$ and $14.5$ km, 
respectively, meaning that SFHx is softer than TM1. 
 Note that SFHx is the best-fit model with the observational mass-radius
 relation of cold NSs \citep{steiner10,SFH}. 
%Both EOSs are compatible with the 2 $M_{\odot}$ NS mass measurements \citep{Demorest10,Antoniadis}.

The 3D hydrodynamic evolution is rather similar between SFHx and TM1, which is 
characterized by the prompt convection phase shortly after bounce
($T_{\rm pb}\lesssim 20$ ms with $T_{\rm pb}$ the postbounce time),
 then the linear (or quiescent) phase ($20 \lesssim T_{\rm pb}\lesssim 140$ ms),
 which is followed by the non-linear phase when the SASI dominates over 
  neutrino-driven convection in the postshock region
($140 \lesssim T_{\rm pb} \lesssim 300$ ms).
The key difference is that the softer EOS (SFHx) makes
 the PNS radius and the shock radius at the shock-stall
more compact than those of TM1. This leads to more stronger activity of the SASI
  for SFHx compared to TM1 (see \cite{KurodaT16ApJL} for more 
details).

%We use three EOSs based on the 
%relativistic-mean-field theory with different
% nuclear interaction treatments, which are DD2 and TM1 of \cite{HS} 
%and SFHx of \citep{SFH}. For SFHx, DD2, and TM1\footnote{The symmetry energy $S$ at nuclear saturation density
%is $S=28.67$, 31.67, and 36.95 MeV, respectively. \citep[e.g.,][]{Fischer14}},
%the maximum gravitational mass ($M_{\rm max}$) and
%the radius ($R$) of cold neutron star (NS) in the vertical part of the mass-radius relationship are
% $M_{\rm max}=2.13$, 2.42, and, 2.21 $M_\odot$
%and $R\sim12$, 13, and, 14.5 km, respectively \citep{Fischer14}.
%SFHx is thus softest followed in order by DD2, and TM1.
%Among the three EOSs, DD2 is constructed in a way that fits well with 
%  nuclear experiments \citep{Lattimer13},
% whereas SFHx is the best fit model with the observational mass-radius relationship 
%\citep{SFH}.
%All EOSs are compatible with the $\sim 2 M_{\odot} $ NS mass measurement 

%We study frequently used solar-metallicity models of a 15 $M_{\odot}$ star \citep{WW95},
% an 11.2 $M_{\odot}$ and a 40 $M_{\odot}$ star of \citet{WHW02}, respectively.
%Our 3D-GR models are named by the progenitor mass with the EOS in parenthesis like
%S15.0(SFHx) which represents the progenitor mass of 15.0 $M_\odot$  and the EOS SFHx are used.%

For extracting GWs, we employ a quadrupole formula 
proposed by \citet{shibata03} for numerical relativity simulations. The transverse and the trace-free gravitational
 field is expressed as 
%\begin{eqnarray}
%\label{eq:hij}
$h_{ij}=(A_+(\theta,\phi)e_++A_\times(\theta,\phi) e_\times)/D \equiv h_+e_+ + 
h_{\times}e_\times$,
%\end{eqnarray}
where $A_{+/\times}(\theta,\phi)$ represent amplitude of orthogonally polarized wave 
components with the emission angle $(\theta,\phi)$,
 $e_{+/\times}$ denote unit polarization
 tensors, $D$ is the distance to the source. The circular polarization of GWs is described by the Stokes parameters 
\citep{seto}, which is expressed by the combination 
of the right-handed ($h_R \equiv (h_+-\mathrm{i}h_{\times})/\sqrt{2}$) 
and left-handed ($h_L \equiv (h_++\mathrm{i}h_{\times})/\sqrt{2}$) polarization
 modes (see Equation (1) of \citet{Hayama16}). We focus on the Stokes 
$V$ parameter, which 
 directly characterizes the asymmetry between the right\mbox{-} and left\mbox{-}handed modes of the circular polarization.

Following \citet{Hayama15},
we perform a coherent network analysis of the four detectors (H, L, V, and K) 
using the {\tt RIDGE} pipeline.\footnote{See \citet{logue12,Gossan16,mukherjee17} 
for representative studies using other pipelines.}
%For the detector noise power spectral density (PSD), we used the design 
For simplicity, a Gaussian, stationary noise is assumed, which is 
produced by generating four independent realization of white noise and 
passing them through the finite impulse response
filters having transfer functions which approximately match the design sensitivity curves of the detectors (taken from  \citet{schutz09,manzotti12,Aso13}).
The injected signals correspond to a signal source located in the direction 
to the Galactic center, where arrival time to each detector is taken into account
 for the source localization with the angular resolution of 
$d\Omega = 4 \times 4$ square degrees (see \citet{Hayama15} for more detail).
%The network data consists of four simulated Gaussian noise with design 
%sensitivities of each detector in which the SHFx signals are injected taking 
%care of arrival time to each detector. 
%The simulated data were generated at a sampling frequency of $4096$ Hz and
%then passed through the data conditioning pipeline. Besides downsampling the 
%data by the factor of $2$, by applying the same anti-aliasing filter to 
%all data streams, the data conditioning pipeline applies whitening filters 
%that were trained on the first $6$ seconds of data for each detector 
%without any injected signal. 
%See \citet{Hayama15} for more detailed procedure of the coherent network analysis.

%\vspace{-0.8cm}
%\begin{equation}
%\begin{aligned}
%&\left (
%\begin{array}{cc}
%  h_R(f,{\hat n})h_R(f',{\hat n}')^* &  h_L(f,{\hat n})h_R(f',{\hat n}')^* \\
%  h_R(f,{\hat n})h_L(f',{\hat n}')^* &  h_L(f,{\hat n})h_L(f',{\hat n}')^*
%\end{array}
%\right )\\
%& = \frac{1}{4\pi} \delta^2_D({\hat n}-{\hat n}')\delta_D(f-f')\\
%&\left (
%\begin{array}{cc}
% I(f,{\hat n})+V(f,{\hat n}) & Q(f,{\hat n})-iU(f,{\hat n}) \\
% Q(f,{\hat n})+iU(f',{\hat n}) & I(f,{\hat n})-V(f,{\hat n})
%\end{array}
%\right ), &
%\end{aligned}
%\end{equation}
%where $f$ is a frequency, ${\hat n}$ is a unit vector in the direction of the propagation, $\delta_D$ is the Dirac's delta function, respectively. Stokes parameter $V$ characterizes the asymmetry between the right\mbox{-} and left\mbox{-}handed modes, whereas parameter $I(\geq|V|)$ corresponds to the total amplitudes.

\section{Results}\label{result}
%\subsubsection{Time-frequency distribution of circular polarization}

 Figure \ref{f1} summarizes the gravitational waveforms (top panels) 
 and the spectrograms of the Stokes $V$ parameter (bottom panels) 
 for SFHx (left panels) and TM1 (right panels), respectively.
 After core bounce,
 one can see a ramp-up feature in the spectrogram of SFHx (the bottom left panel)
 before $T_{\rm pb} \sim 140$ ms,  where the excess (colored by blue and red) 
increases from $\sim 100$ to 500 Hz. 
This ramp-up component is clearly correlated with the peak 
 frequency of the PNS surface oscillation (black line in the bottom
 panels, corresponding to $F_{\rm peak}$ defined in \citet{BMuller13}). Note that 
we focus on the frequency range below $\sim$ 500 Hz in the spectrogram,
 because the higher frequency domain 
is difficult to detect due to a shot-noise of the 
laser interferometers. The ramp-up feature is also seen for TM1 (the bottom
 right panel) for a longer period ($T_{\rm pb} \lesssim 180$ ms). The PNS 
 surface oscillation, primarily driven by buoyancy \citep{Murphy09}, 
 occurs in a stochastic manner. Consequently, 
 the sign of the circular polarization changes stochastically. 
In fact, the ramp-up signature of the 
$V$ parameter ($T_{\rm pb} \lesssim 180$ ms in the bottom panels) is 
a mixture of right-handed (colored by red) and 
left-handed (colored by blue) modes, where
the right-handed mode is by chance stronger for SFHx.

At $140 \lesssim T_{\rm pb} \lesssim 300$ ms, the waveform of SFHx shows a 
quasi-periodic modulation (the top left panel), which is not present
in the waveform of TM1 (the top right panel). As previously identified, 
this signature is a result of the SASI-induced mass accretion flows striking 
the PNS core surface (e.g., \citet{KurodaT16ApJL,Andresen17}).
In the SASI-dominant phase, the $V$-mode spectrogram for SFHx (the bottom 
left panel) shows a clear excess of the right-handed polarization 
first (colored by red, $140 \lesssim T_{\rm pb} \lesssim 200$ ms), which is
followed by a clear excess of the left-handed polarization 
(colored by blue, $220 \lesssim T_{\rm pb} \lesssim 320$ ms). 
 Between the two epochs, a quiescent phase with vanishing polarization amplitude
 (white in the panel) is observed between $200 \lesssim T_{\rm pb} 
\lesssim 220$ ms. 

\begin{figure}
\begin{center}
\includegraphics[width=1.0\linewidth]{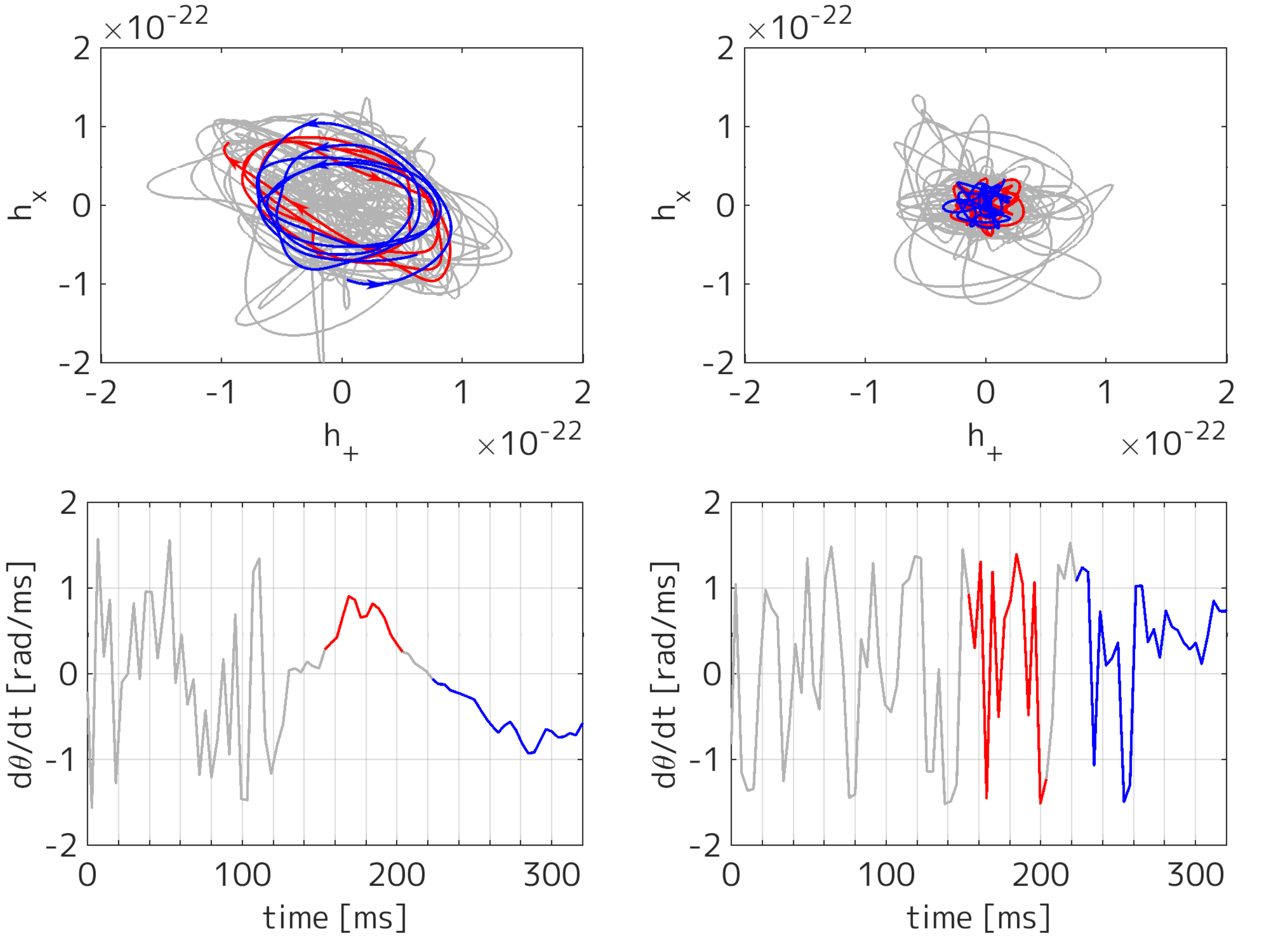}
\caption{The top panels show the trajectory of the GW polarization on the 
$h_{+}-h_{\times}$ plane of SFHx (left panel) and TM1 (right panel). 
  Making a correspondence to the bottom left panel
 of Figure \ref{f1}, the two characteristic epoch with the right-handed 
($140 \lesssim T_{\rm pb} \lesssim 200$ ms) and left-handed 
($220 \lesssim T_{\rm pb} \lesssim 320$ ms) polarization are highlighted with 
the red and blue color. The gray line denotes the whole 
trajectory during the simulation time. Similar to the top panels, but 
the bottom panels depict the time derivative of the polarization angle ($d\theta/dt$).
 The polarization angle $\theta$ is measured from the axis of $h_{\times}$ 
as $\theta \equiv \arctan (h_{+}/h_{\times})$ spanning $0 \leq \theta \leq \pi$  and $ 0 \geq \theta \geq -\pi$, respectively.
Note that the signals are low-pass filtered with a cutoff frequency of $400$Hz
 to focus on the low-frequency behaviors.
%Note that the time resolution of $1.95$ms. 
}
\label{f2}
\end{center}
\end{figure}

 Figure \ref{f2} visualizes the time evolution of the GW circular 
polarization of SFHx (left panels) and TM1 (right panels).
 Note in each panel that the two SASI-dominant phases of SFHx
 are colored by red or blue, corresponding to the bottom left panel 
of Figure \ref{f1} at $T_{\rm pb} \gtrsim 140$ ms.
 One can see a clearer
 polarization signature for SFHx (the top left panel) characterized by the bigger
 GW amplitude with the right-handed (red line) and left-handed mode (blue line)
 than those for TM1 (the top right panel).
 Before $T_{\rm pb} \sim 140$ ms, the bottom left panel
 shows that the time derivative of the polarization angle ($d\theta/dt$)
 changes randomly with time, taking both positive and negative values of $d\theta/dt$. 
But, after $T_{\rm pb} \sim 140$ ms when the 
SASI activity begins to be vigorous (e.g., \citet{KurodaT16ApJL}), the 
right-handed GW polarization (e.g., the red line in the bottom left panel)
 is shown to transit to the left-handed polarization 
(the blue line) until $T_{\rm pb} \sim 320$ ms, after which
neutrino-driven convection dominates over the SASI. The bottom right 
 panel shows that the value of $d\theta/dt$ of TM1 changes more 
stochastically during 
 the simulation time, making the net circular 
 polarization very small (e.g., the top right panel of Figure \ref{f2} and 
bottom right panel of Figure \ref{f1}).

%Figure 2 shows angular evolution between $h_+$ and $h_\times$. 
% The left and right panels are 
%for SFHx and TM1, respectively. 
%The top left panel corresponds to the $h_+$-$h_\times$ trajectory. 
%The red and blue lines with arrows highlight the regions where the $V$-mode 
%with the central frequency of $\sim 100$Hz ramps up. 
%The bottom left panel corresponds to the derivative of angle 
%$\theta (t)$ between $h_+$ and $h_\times$ with the direction to $h_+$ from 
%$h_\times$ with respect to time $t$. 
%The red and blue lines correspond to the same region as the above panels. 
%Here the angle $theta(t)$, where the values are in the closed interval 
%$[-\pi,\pi]$, is calculated from the four-quadrant inverse tangent. 
%In first $150$ms from the core-bounce time, the direction of rotation 
%is not oriented clearly. After $150$ms, there is strong orientation 
%of the rotation. In $150$-$220$ms and $220$-$320$ms, right-handed and 
%left-handed direction of the rotation are exceeded, respectively, 
%which are consistent with the excess of the $V$-mode in $\sim 100$Hz 
%in the figure 1. This suggests the axis of the rotation of the SASI changed 
%around $220$ms. The top and bottom right panels correspond to the 
%$h_+$\mbox{-}$h_\times$ trajectory and $d\theta(t)/dt$ of TM1.  
%In this model no clear orientation of $d\theta(t)/dt$ appear in the
% corresponding time regions. 

From the spectrogram of SFHx (the bottom left panel of Figure \ref{f1}),
 one can clearly see that the typical frequency of the strong circular 
polarization is in the range of 100 $\sim$ 200 Hz, which is in the best sensitivity 
range of the LIGO-class detectors. We thus focus on the GW signatures of 
SFHx in the following.

\begin{figure}
\begin{center}
\includegraphics[width=0.7\linewidth]{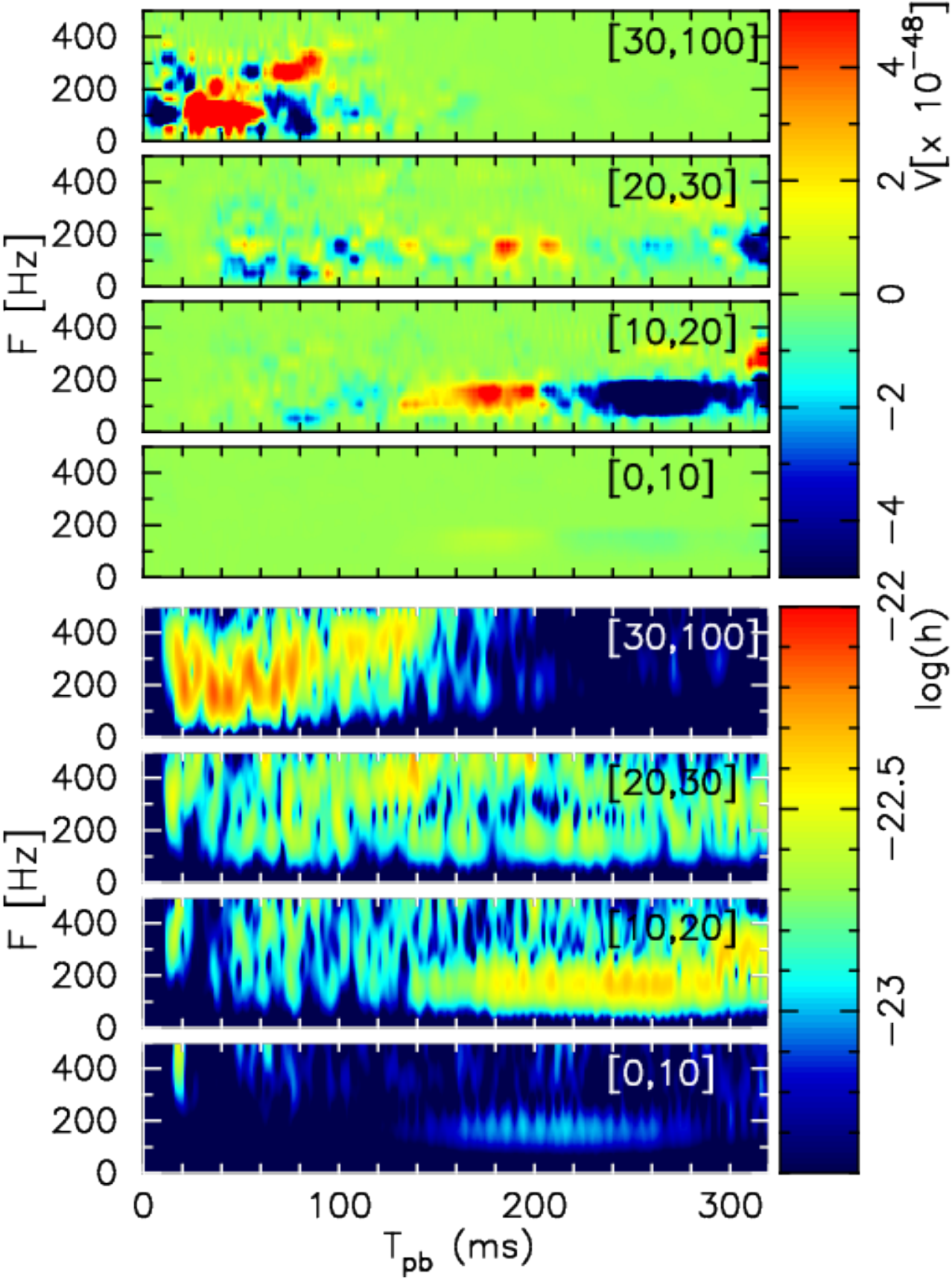}
\caption{Contributions from four spherical shells with intervals of [0, 10], [10,
20], [20, 30], and [30, 100] km to the GW circular polarization 
(top panels) and the GW amplitude (bottom panels) of SFHx, respectively. The source is assumed 
at a distance of 10 kpc.}
\label{f3}
\end{center}
\end{figure}

To clarify the origin of the circular polarization in the postbounce 
core, 
we plot in Figure \ref{f3} contributions from representative 
four spherical shells to the circular polarization (the top panels) and 
 the GW amplitude (the bottom panels) of SFHx, respectively.
 From the PNS core surface region (see the panels labeled with
 [10,20] km), the excess in the circular polarization (see the horizontal
 stripe colored in red or blue in the top panel) 
is correlated with the GW emission
 at the low-frequency domain of 100 - 200 Hz (colored by yellow or red in 
the bottom panel). This low-frequency  component is identified 
to originate from the SASI-induced flows penetrating into the PNS core 
\citep{KurodaT16ApJL,Andresen17}. From between the PNS and the shock 
(see panels labeled with $\sim$ [30,100] km), 
  one can also see that the ramp-up $g-$mode of the PNS oscillation
 (e.g., the red and yellow region in the bottom panel at $0 \lesssim T_{\rm pb} 
\lesssim 140$ ms) 
has a significant overlap with the excess in the $V$ parameter 
(the top panel). These findings illuminate the importance of detecting the GW polarization 
 because the pre-explosion hydrodyamical features such as the 
SASI activity and the PNS oscillation are clearly imprinted in the 
spectrogram.
%The signal data is divided into $20$ms long segments. One sample is shifted 
%to take next segment. The Stokes parameters are calculated for each segment. 
%The interval of frequency is $8$Hz via interpolation of the Stokes parameters. 
%In the left panel it is seen that the $V$-mode around $100$Hz in $150$-$320$ms 
%is ramped up in the $[10,20]$km shell which is the same region as the one where
% the SASI activities are excited. This suggests the $V$-mode is generated by 
%the SASI activities. 

%The detector network used here 
%consists of aLIGO Hanford(H), aLIGO Livingston(L), adVirgo(V), and KAGRA(K).
% For the detector noise power spectral density (PSD), we used the design 
%sensitivity curves for aLIGO, adVirgo, and KAGRA as given in 
%\citet{schutz09,manzotti12}, and kept the locations and orientation the same as 
%the real detectors. 
%$50$ second long Gaussian, stationary noise was generated 
%by first generating $4$ independent realizations of white noise and then 
%passing them through finite impulse regression (FIR) filters having transfer 
%functions that approximately match the design curves. The SFHx signals of 
%constant amplitude were added to the simulated noise every $3$ seconds.
%\subsubsection{Detectability of the circular polarization}
\begin{figure*}
\begin{center}
\includegraphics[width=12cm]{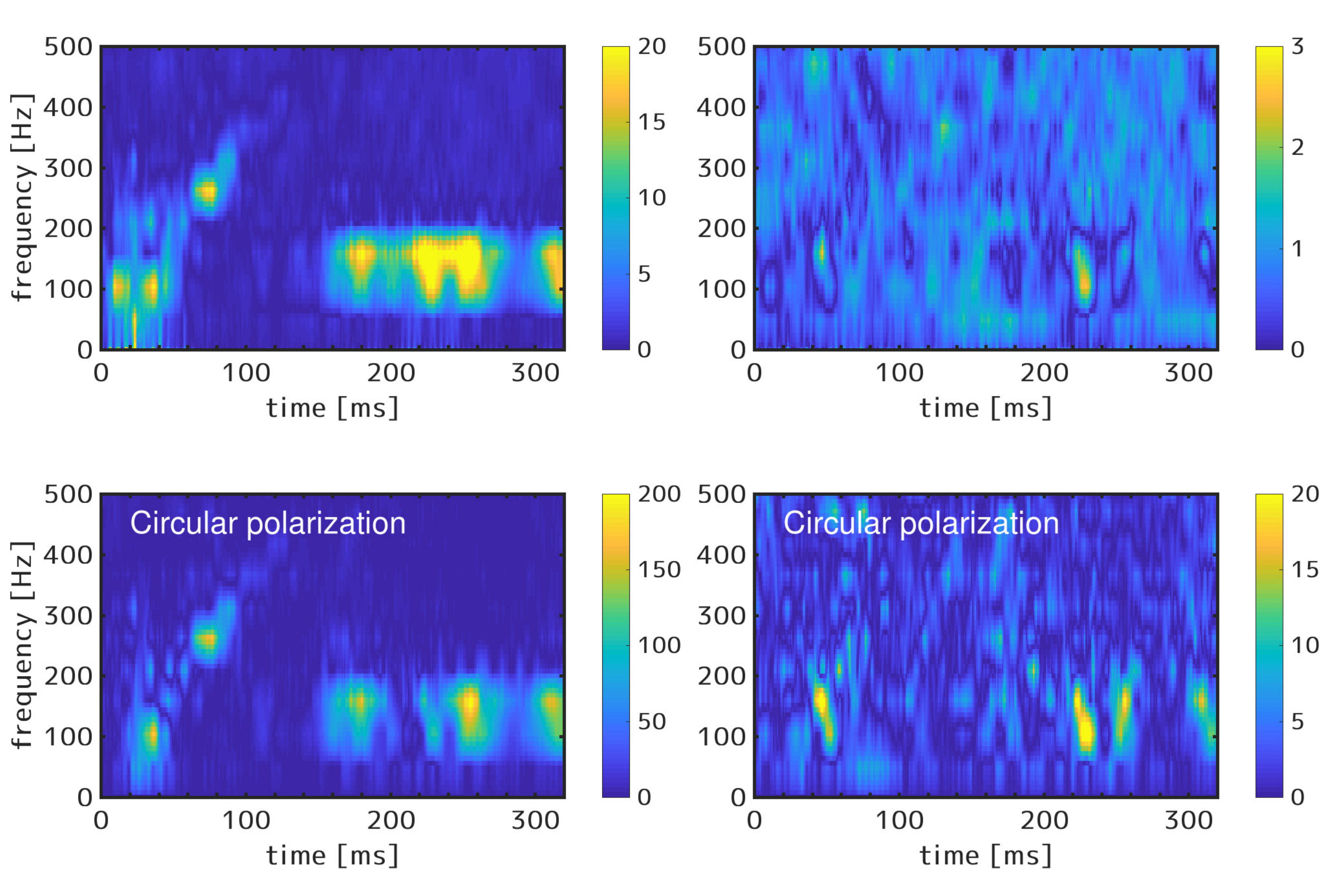}
\caption{Time-frequency signal-to-noise ratio of the reconstructed waveform
 (${\rm SNR}_{\rm TF}$, top panels) and the circular polarization 
(${\rm SNR}_{\rm CP}$, bottom panels) for a source at 2 kpc (left panels) and 
 at 10 kpc (right panels), respectively (see text).
 Note that the excess changes slightly with the distance because the noise
 impacts non-linearly the phase between $h_{+}$ and $h_{\times}$.}
\label{f4}
\end{center}
\end{figure*}

Following \citet{Hayama15}, we quantify 
the detectability of the GW signatures by performing 
Monte Carlo simulations, where the network of H, L, V, and K is considered.
The $V$-mode is calculated by the reconstructed $h_+$ and $h_{\times}$.
The top panels of Figure \ref{f4} shows a {\it time-frequency}
 signal-to-noise ratio (${\rm SNR}_{\rm TF}$) of the reconstructed GW waveform, which
 is defined as ${\rm SNR}_{\rm TF} \equiv I(s + n)
/\sqrt{\langle I(n) - \langle I(n) \rangle \rangle^2}$  with $I(x)$ the
 Stokes $I$ parameter (e.g., Eq.(1) in \citet{Hayama16}), $s$ the signal, and $n$ the Gaussian noise, $\langle x \rangle$
 denotes the time average of $x$, respectively. 
By setting the (optimal) detection threshold as 10, 
the top panels of Figure \ref{f4} show that
 the detection horizon even using the four detectors 
can only reach to a several kpc (e.g., the top left panel 
for a source at 2kpc).  
By replacing $I$ of ${\rm SNR}_{\rm TF}$ with $|V|$, we 
estimate the SNR of the circular polarization 
(${\rm SNR}_{\rm CP}$).
From the bottom panels of Figure \ref{f4}, one
 can see that the SNR of the circular polarization is significantly higher
 than ${\rm SNR}_{\rm TF}$. 
For a source at 10 kpc (the bottom right panel), ${\rm SNR}_{\rm CP}$ exceeds
 the fiducial threshold for the low-frequency component, which cannot 
be detectable solely by looking for the excess in the GW spectrogram 
(the top right panel). The significant enhancement of ${\rm SNR}_{\rm CP}$
comparing with ${\rm SNR}_{\rm TF}$ is because the Gaussian noise (by nature)
 has little component of the circular polarization. From the 
 bottom left panel, one can see that the circular 
polarization from the ramp-up component of the PNS oscillation 
is high (${\rm SNR}_{\rm CP} \sim 200$), which may allow detection 
for a nearby source such as at 2 kpc (the bottom left panel).
 These results demonstrate that the 
GW circular polarization could not only extend the detection horizon of the 
 signal farther% comparing to the GW search solely relying on the wave 
%amplitude
, but also provide a new probe to 
decipher the inner-working of the supernova engine such as
 the PNS oscillation and the SASI.

\section{Discussions}
In this work, we considered only the idealized Gaussian noise. Effects of
real non-Gaussian and nonstationary noise need to be considered
 \citep{Powell16,Powell17}. This is one of the most important tasks that we have to 
investigate as a sequel of this work. It is recently pointed out that there is 
a correlation between the SASI-induced modulation in the GW and neutrino signals \citep{kuroda17,Tamborra13}.
 The correlation between the neutrino signal and the GW circular polarization 
 would deserve further investigation.
% One could in principle do this, 
%but only if one could afford enough computational time to make the many 3D CCSN 
%runs doable
 Using a single progenitor model and the particular EOS (SFHx), we 
found that the strong SASI activity and the PNS oscillation 
are imprinted in both the waveform and the circular polarization. In order 
 to clarify how general this trend would be, a systematic 3D CCSN simulation
 changing the progenitor models and EOSs is needed to be done. This is a grand 
computational challenge, which we shall leave as future work.
 Consideration of the circular polarization 
into the GW search pipeline is also urgent, which could
 significantly enhance the chance of detecting a CCSN GW in the future.

 \section*{Acknowledgements}
 We are thankful to S. Yamada, H. Kawahara and N. Kanda for stimulating discussions.
Numerical computations were carried out in part on 
XC30 at the CfCA of NAOJ.
This study was supported by JSPS KAKENHI Grant Number
 (JP15H00789, JP15H01039, JP17H05206, JP17K14306, JP17H01130, JP17H06364), and by the Central 
Research Institute of Fukuoka University (Nos.171042, 177103), 
and JICFuS as a priority issue to be tackled by using
 the Post `K' Computer.

%%%%%%%%%%%%%%%%%%%%%%%%%%%%%%%%%%%%%%%%%%%%%%%%%%


\vspace{-0.5cm}
\bibliographystyle{mnras}
\bibliography{mybib_hayama} % if your bibtex file is called example.bib

\begin{thebibliography}{}
\makeatletter
\relax
\def\mn@urlcharsother{\let\do\@makeother \do\$\do\&\do\#\do\^\do\_\do\%\do\~}
\def\mn@doi{\begingroup\mn@urlcharsother \@ifnextchar [ {\mn@doi@}
  {\mn@doi@[]}}
\def\mn@doi@[#1]#2{\def\@tempa{#1}\ifx\@tempa\@empty \href
  {http://dx.doi.org/#2} {doi:#2}\else \href {http://dx.doi.org/#2} {#1}\fi
  \endgroup}
\def\mn@eprint#1#2{\mn@eprint@#1:#2::\@nil}
\def\mn@eprint@arXiv#1{\href {http://arxiv.org/abs/#1} {{\tt arXiv:#1}}}
\def\mn@eprint@dblp#1{\href {http://dblp.uni-trier.de/rec/bibtex/#1.xml}
  {dblp:#1}}
\def\mn@eprint@#1:#2:#3:#4\@nil{\def\@tempa {#1}\def\@tempb {#2}\def\@tempc
  {#3}\ifx \@tempc \@empty \let \@tempc \@tempb \let \@tempb \@tempa \fi \ifx
  \@tempb \@empty \def\@tempb {arXiv}\fi \@ifundefined
  {mn@eprint@\@tempb}{\@tempb:\@tempc}{\expandafter \expandafter \csname
  mn@eprint@\@tempb\endcsname \expandafter{\@tempc}}}

\bibitem[\protect\citeauthoryear{{Abbott} et~al.,}{{Abbott}
  et~al.}{2016}]{GW_review}
{Abbott} B.~P.,  et~al., 2016, \mn@doi [Living Reviews in Relativity]
  {10.1007/lrr-2016-1}, \href {http://ads.nao.ac.jp/abs/2016LRR....19....1A}
  {19, 1}

\bibitem[\protect\citeauthoryear{{Abbott} et~al.,}{{Abbott}
  et~al.}{2017}]{GW2_virgo}
{Abbott} B.~P.,  et~al., 2017, \mn@doi [Physical Review Letters]
  {10.1103/PhysRevLett.119.141101}, \href
  {http://ads.nao.ac.jp/abs/2017PhRvL.119n1101A} {119, 141101}

\bibitem[\protect\citeauthoryear{{Akutsu} et~al.,}{{Akutsu}
  et~al.}{2017}]{kagra17}
{Akutsu} T.,  et~al., 2017, preprint, \href
  {http://ads.nao.ac.jp/abs/2017arXiv171200148A} {} (\mn@eprint {arXiv}
  {1712.00148})

\bibitem[\protect\citeauthoryear{{Andresen}, {M{\"u}ller}, {M{\"u}ller}  \&
  {Janka}}{{Andresen} et~al.}{2017}]{Andresen17}
{Andresen} H.,  {M{\"u}ller} B.,  {M{\"u}ller} E.,   {Janka} H.-T.,  2017,
  \mn@doi [\mnras] {10.1093/mnras/stx618}, \href
  {http://cdsads.u-strasbg.fr/abs/2017MNRAS.468.2032A} {468, 2032}

\bibitem[\protect\citeauthoryear{{Aso}, {Michimura}, {Somiya}, {Ando},
  {Miyakawa}, {Sekiguchi}, {Tatsumi}  \& {Yamamoto}}{{Aso}
  et~al.}{2013}]{Aso13}
{Aso} Y.,  {Michimura} Y.,  {Somiya} K.,  {Ando} M.,  {Miyakawa} O.,
  {Sekiguchi} T.,  {Tatsumi} D.,   {Yamamoto} H.,  2013, \mn@doi [\prd]
  {10.1103/PhysRevD.88.043007}, \href
  {http://cdsads.u-strasbg.fr/abs/2013PhRvD..88d3007A} {88, 043007}

\bibitem[\protect\citeauthoryear{{Baumgarte} \& {Shapiro}}{{Baumgarte} \&
  {Shapiro}}{1999}]{Baumgarte99}
{Baumgarte} T.~W.,  {Shapiro} S.~L.,  1999, \mn@doi [\prd]
  {10.1103/PhysRevD.59.024007}, \href
  {http://ads.nao.ac.jp/abs/1999PhRvD..59b4007B} {59, 024007}

\bibitem[\protect\citeauthoryear{{Cerd{\'a}-Dur{\'a}n}, {DeBrye}, {Aloy},
  {Font}  \& {Obergaulinger}}{{Cerd{\'a}-Dur{\'a}n}
  et~al.}{2013}]{CerdaDuran13}
{Cerd{\'a}-Dur{\'a}n} P.,  {DeBrye} N.,  {Aloy} M.~A.,  {Font} J.~A.,
  {Obergaulinger} M.,  2013, \mn@doi [\apjl] {10.1088/2041-8205/779/2/L18},
  \href {http://cdsads.u-strasbg.fr/abs/2013ApJ...779L..18C} {779, L18}

\bibitem[\protect\citeauthoryear{{Gossan}, {Sutton}, {Stuver}, {Zanolin},
  {Gill}  \& {Ott}}{{Gossan} et~al.}{2016}]{Gossan16}
{Gossan} S.~E.,  {Sutton} P.,  {Stuver} A.,  {Zanolin} M.,  {Gill} K.,   {Ott}
  C.~D.,  2016, \mn@doi [\prd] {10.1103/PhysRevD.93.042002}, \href
  {http://cdsads.u-strasbg.fr/abs/2016PhRvD..93d2002G} {93, 042002}

\bibitem[\protect\citeauthoryear{{Hayama}, {Kuroda}, {Kotake}  \&
  {Takiwaki}}{{Hayama} et~al.}{2015}]{Hayama15}
{Hayama} K.,  {Kuroda} T.,  {Kotake} K.,   {Takiwaki} T.,  2015, \mn@doi [\prd]
  {10.1103/PhysRevD.92.122001}, \href
  {http://ads.nao.ac.jp/abs/2015PhRvD..92l2001H} {92, 122001}

\bibitem[\protect\citeauthoryear{{Hayama}, {Kuroda}, {Nakamura}  \&
  {Yamada}}{{Hayama} et~al.}{2016}]{Hayama16}
{Hayama} K.,  {Kuroda} T.,  {Nakamura} K.,   {Yamada} S.,  2016, \mn@doi
  [Physical Review Letters] {10.1103/PhysRevLett.116.151102}, \href
  {http://adsabs.harvard.edu/abs/2016PhRvL.116o1102H} {116, 151102}

\bibitem[\protect\citeauthoryear{{Heger}, {Woosley}  \& {Spruit}}{{Heger}
  et~al.}{2005}]{Heger05}
{Heger} A.,  {Woosley} S.~E.,   {Spruit} H.~C.,  2005, \mn@doi [\apj]
  {10.1086/429868}, \href {http://ads.nao.ac.jp/abs/2005ApJ...626..350H} {626,
  350}

\bibitem[\protect\citeauthoryear{{Hempel} \& {Schaffner-Bielich}}{{Hempel} \&
  {Schaffner-Bielich}}{2010}]{HS}
{Hempel} M.,  {Schaffner-Bielich} J.,  2010, \mn@doi [Nuclear Physics A]
  {10.1016/j.nuclphysa.2010.02.010}, \href
  {http://adsabs.harvard.edu/abs/2010NuPhA.837..210H} {837, 210}

\bibitem[\protect\citeauthoryear{{Janka}}{{Janka}}{2017}]{janka17}
{Janka} H.-T.,  2017, preprint, \href
  {http://ads.nao.ac.jp/abs/2017arXiv170208825J} {} (\mn@eprint {arXiv}
  {1702.08825})

\bibitem[\protect\citeauthoryear{{Kotake}}{{Kotake}}{2013}]{Kotake13}
{Kotake} K.,  2013, \mn@doi [Comptes Rendus Physique]
  {10.1016/j.crhy.2013.01.008}, \href
  {http://adsabs.harvard.edu/abs/2013CRPhy..14..318K} {14, 318}

\bibitem[\protect\citeauthoryear{{Kotake}, {Iwakami}, {Ohnishi}  \&
  {Yamada}}{{Kotake} et~al.}{2009}]{Kotake09}
{Kotake} K.,  {Iwakami} W.,  {Ohnishi} N.,   {Yamada} S.,  2009, \mn@doi
  [\apjl] {10.1088/0004-637X/697/2/L133}, \href
  {http://ads.nao.ac.jp/abs/2009ApJ...697L.133K} {697, L133}

\bibitem[\protect\citeauthoryear{{Kuroda}, {Kotake}  \& {Takiwaki}}{{Kuroda}
  et~al.}{2012}]{KurodaT12}
{Kuroda} T.,  {Kotake} K.,   {Takiwaki} T.,  2012, \mn@doi [\apj]
  {10.1088/0004-637X/755/1/11}, \href
  {http://ads.nao.ac.jp/abs/2012ApJ...755...11K} {755, 11}

\bibitem[\protect\citeauthoryear{{Kuroda}, {Takiwaki}  \& {Kotake}}{{Kuroda}
  et~al.}{2014}]{KurodaT14}
{Kuroda} T.,  {Takiwaki} T.,   {Kotake} K.,  2014, \mn@doi [\prd]
  {10.1103/PhysRevD.89.044011}, \href
  {http://ads.nao.ac.jp/abs/2014PhRvD..89d4011K} {89, 044011}

\bibitem[\protect\citeauthoryear{{Kuroda}, {Kotake}  \& {Takiwaki}}{{Kuroda}
  et~al.}{2016}]{KurodaT16ApJL}
{Kuroda} T.,  {Kotake} K.,   {Takiwaki} T.,  2016, \mn@doi [\apjl]
  {10.3847/2041-8205/829/1/L14}, \href
  {http://ads.nao.ac.jp/abs/2016ApJ...829L..14K} {829, L14}

\bibitem[\protect\citeauthoryear{{Kuroda}, {Kotake}, {Hayama}  \&
  {Takiwaki}}{{Kuroda} et~al.}{2017}]{kuroda17}
{Kuroda} T.,  {Kotake} K.,  {Hayama} K.,   {Takiwaki} T.,  2017, \mn@doi [\apj]
  {10.3847/1538-4357/aa988d}, \href
  {http://ads.nao.ac.jp/abs/2017ApJ...851...62K} {851, 62}

\bibitem[\protect\citeauthoryear{{Logue}, {Ott}, {Heng}, {Kalmus}  \&
  {Scargill}}{{Logue} et~al.}{2012}]{logue12}
{Logue} J.,  {Ott} C.~D.,  {Heng} I.~S.,  {Kalmus} P.,   {Scargill} J.~H.~C.,
  2012, \mn@doi [\prd] {10.1103/PhysRevD.86.044023}, \href
  {http://ads.nao.ac.jp/abs/2012PhRvD..86d4023L} {86, 044023}

\bibitem[\protect\citeauthoryear{{Manzotti} \& {Dietz}}{{Manzotti} \&
  {Dietz}}{2012}]{manzotti12}
{Manzotti} A.,  {Dietz} A.,  2012, preprint, \href
  {http://adsabs.harvard.edu/abs/2012arXiv1202.4031M} {} (\mn@eprint {arXiv}
  {1202.4031})

\bibitem[\protect\citeauthoryear{{Morozova}, {Radice}, {Burrows}  \&
  {Vartanyan}}{{Morozova} et~al.}{2018}]{viktoriya18}
{Morozova} V.,  {Radice} D.,  {Burrows} A.,   {Vartanyan} D.,  2018, preprint,
  \href {http://ads.nao.ac.jp/abs/2018arXiv180101914M} {} (\mn@eprint {arXiv}
  {1801.01914})

\bibitem[\protect\citeauthoryear{{Mukherjee}, {Salazar}, {Mittelstaedt}  \&
  {Valdez}}{{Mukherjee} et~al.}{2017}]{mukherjee17}
{Mukherjee} S.,  {Salazar} L.,  {Mittelstaedt} J.,   {Valdez} O.,  2017,
  \mn@doi [\prd] {10.1103/PhysRevD.96.104033}, \href
  {http://adsabs.harvard.edu/abs/2017PhRvD..96j4033M} {96, 104033}

\bibitem[\protect\citeauthoryear{{M\"uller} \& {Janka}}{{M\"uller} \&
  {Janka}}{1997}]{EMuller97}
{M\"uller} E.,  {Janka} H.-T.,  1997, \aap, \href
  {http://ads.nao.ac.jp/abs/1997A%26A...317..140M} {317, 140}

\bibitem[\protect\citeauthoryear{{M{\"u}ller}, {Rampp}, {Buras}, {Janka}  \&
  {Shoemaker}}{{M{\"u}ller} et~al.}{2004}]{EMuller04}
{M{\"u}ller} E.,  {Rampp} M.,  {Buras} R.,  {Janka} H.-T.,   {Shoemaker} D.~H.,
   2004, \mn@doi [\apj] {10.1086/381360}, \href
  {http://ads.nao.ac.jp/abs/2004ApJ...603..221M} {603, 221}

\bibitem[\protect\citeauthoryear{{M{\"u}ller}, {Janka}  \&
  {Marek}}{{M{\"u}ller} et~al.}{2013}]{BMuller13}
{M{\"u}ller} B.,  {Janka} H.-T.,   {Marek} A.,  2013, \mn@doi [\apj]
  {10.1088/0004-637X/766/1/43}, \href
  {http://ads.nao.ac.jp/abs/2013ApJ...766...43M} {766, 43}

\bibitem[\protect\citeauthoryear{{Murphy}, {Ott}  \& {Burrows}}{{Murphy}
  et~al.}{2009}]{Murphy09}
{Murphy} J.~W.,  {Ott} C.~D.,   {Burrows} A.,  2009, \mn@doi [\apj]
  {10.1088/0004-637X/707/2/1173}, \href
  {http://ads.nao.ac.jp/abs/2009ApJ...707.1173M} {707, 1173}

\bibitem[\protect\citeauthoryear{{Ott}}{{Ott}}{2009}]{Ott09}
{Ott} C.~D.,  2009, \mn@doi [Classical and Quantum Gravity]
  {10.1088/0264-9381/26/6/063001}, \href
  {http://ads.nao.ac.jp/abs/2009CQGra..26f3001O} {26, 063001}

\bibitem[\protect\citeauthoryear{{Ott} et~al.,}{{Ott} et~al.}{2013}]{Ott13}
{Ott} C.~D.,  et~al., 2013, \mn@doi [\apj] {10.1088/0004-637X/768/2/115}, \href
  {http://cdsads.u-strasbg.fr/abs/2013ApJ...768..115O} {768, 115}

\bibitem[\protect\citeauthoryear{{Powell}, {Gossan}, {Logue}  \&
  {Heng}}{{Powell} et~al.}{2016}]{Powell16}
{Powell} J.,  {Gossan} S.~E.,  {Logue} J.,   {Heng} I.~S.,  2016, \mn@doi
  [\prd] {10.1103/PhysRevD.94.123012}, \href
  {http://cdsads.u-strasbg.fr/abs/2016PhRvD..94l3012P} {94, 123012}

\bibitem[\protect\citeauthoryear{{Powell}, {Szczepanczyk}  \& {Heng}}{{Powell}
  et~al.}{2017}]{Powell17}
{Powell} J.,  {Szczepanczyk} M.,   {Heng} I.~S.,  2017, preprint, \href
  {http://adsabs.harvard.edu/abs/2017arXiv170900955P} {} (\mn@eprint {arXiv}
  {1709.00955})

\bibitem[\protect\citeauthoryear{{Sathyaprakash} \& {Schutz}}{{Sathyaprakash}
  \& {Schutz}}{2009}]{schutz09}
{Sathyaprakash} B.~S.,  {Schutz} B.~F.,  2009, \mn@doi [Living Reviews in
  Relativity] {10.12942/lrr-2009-2}, \href
  {http://adsabs.harvard.edu/abs/2009LRR....12....2S} {12, 2}

\bibitem[\protect\citeauthoryear{{Seto} \& {Taruya}}{{Seto} \&
  {Taruya}}{2007}]{seto}
{Seto} N.,  {Taruya} A.,  2007, \mn@doi [Physical Review Letters]
  {10.1103/PhysRevLett.99.121101}, \href
  {http://adsabs.harvard.edu/abs/2007PhRvL..99l1101S} {99, 121101}

\bibitem[\protect\citeauthoryear{{Shibata} \& {Nakamura}}{{Shibata} \&
  {Nakamura}}{1995}]{Shibata95}
{Shibata} M.,  {Nakamura} T.,  1995, \mn@doi [\prd] {10.1103/PhysRevD.52.5428},
  \href {http://ads.nao.ac.jp/abs/1995PhRvD..52.5428S} {52, 5428}

\bibitem[\protect\citeauthoryear{{Shibata} \& {Sekiguchi}}{{Shibata} \&
  {Sekiguchi}}{2003}]{shibata03}
{Shibata} M.,  {Sekiguchi} Y.-I.,  2003, \mn@doi [\prd]
  {10.1103/PhysRevD.68.104020}, \href
  {http://ads.nao.ac.jp/abs/2003PhRvD..68j4020S} {68, 104020}

\bibitem[\protect\citeauthoryear{{Steiner}, {Lattimer}  \& {Brown}}{{Steiner}
  et~al.}{2010}]{steiner10}
{Steiner} A.~W.,  {Lattimer} J.~M.,   {Brown} E.~F.,  2010, \mn@doi [\apj]
  {10.1088/0004-637X/722/1/33}, \href
  {http://ads.nao.ac.jp/abs/2010ApJ...722...33S} {722, 33}

\bibitem[\protect\citeauthoryear{{Steiner}, {Hempel}  \& {Fischer}}{{Steiner}
  et~al.}{2013}]{SFH}
{Steiner} A.~W.,  {Hempel} M.,   {Fischer} T.,  2013, \mn@doi [\apj]
  {10.1088/0004-637X/774/1/17}, \href
  {http://adsabs.harvard.edu/abs/2013ApJ...774...17S} {774, 17}

\bibitem[\protect\citeauthoryear{{Summa}, {Janka}, {Melson}  \&
  {Marek}}{{Summa} et~al.}{2017}]{summa17}
{Summa} A.,  {Janka} H.-T.,  {Melson} T.,   {Marek} A.,  2017, preprint, \href
  {http://ads.nao.ac.jp/abs/2017arXiv170804154S} {} (\mn@eprint {arXiv}
  {1708.04154})

\bibitem[\protect\citeauthoryear{{Takiwaki}, {Kotake}  \& {Suwa}}{{Takiwaki}
  et~al.}{2016}]{takiwaki16}
{Takiwaki} T.,  {Kotake} K.,   {Suwa} Y.,  2016, \mn@doi [\mnras]
  {10.1093/mnrasl/slw105}, \href {http://ads.nao.ac.jp/abs/2016MNRAS.461L.112T}
  {461, L112}

\bibitem[\protect\citeauthoryear{{Tamborra}, {Hanke}, {M{\"u}ller}, {Janka}  \&
  {Raffelt}}{{Tamborra} et~al.}{2013}]{Tamborra13}
{Tamborra} I.,  {Hanke} F.,  {M{\"u}ller} B.,  {Janka} H.-T.,   {Raffelt} G.,
  2013, \mn@doi [Physical Review Letters] {10.1103/PhysRevLett.111.121104},
  \href {http://cdsads.u-strasbg.fr/abs/2013PhRvL.111l1104T} {111, 121104}

\bibitem[\protect\citeauthoryear{{Woosley} \& {Weaver}}{{Woosley} \&
  {Weaver}}{1995}]{WW95}
{Woosley} S.~E.,  {Weaver} T.~A.,  1995, \mn@doi [\apjs] {10.1086/192237},
  \href {http://ads.nao.ac.jp/abs/1995ApJS..101..181W} {101, 181}

\bibitem[\protect\citeauthoryear{{Yakunin} et~al.,}{{Yakunin}
  et~al.}{2015}]{Yakunin15}
{Yakunin} K.~N.,  et~al., 2015, \mn@doi [\prd] {10.1103/PhysRevD.92.084040},
  \href {http://cdsads.u-strasbg.fr/abs/2015PhRvD..92h4040Y} {92, 084040}

\bibitem[\protect\citeauthoryear{{Yakunin} et~al.,}{{Yakunin}
  et~al.}{2017}]{Yakunin17}
{Yakunin} K.~N.,  et~al., 2017, preprint, \href
  {http://cdsads.u-strasbg.fr/abs/2017arXiv170107325Y} {} (\mn@eprint {arXiv}
  {1701.07325})

\makeatother
\end{thebibliography}

%%%%%%%%%%%%%%%%%%%%%%%%%%%%%%%%%%%%%%%%%%%%%%%%%%

% Don't change these lines
%\bsp	% typesetting comment
\label{lastpage}
\end{document}